\documentclass[%
 aip,
 amsmath,amssymb,
 reprint,%
]{revtex4-2}

\usepackage{graphicx}
\usepackage{dcolumn}
\usepackage{bm}

\usepackage[utf8]{inputenc}
\usepackage[T1]{fontenc}
\usepackage{mathptmx}
\usepackage{etoolbox}
\usepackage{xcolor}

\makeatletter
\def\@email#1#2{%
 \endgroup
 \patchcmd{\titleblock@produce}
  {\frontmatter@RRAPformat}
  {\frontmatter@RRAPformat{\produce@RRAP{*#1\href{mailto:#2}{#2}}}\frontmatter@RRAPformat}
  {}{}
}%
\makeatother
\begin{document}

\preprint{AIP/123-QED}

\title[S. Mohanty et. al.]{Magnetization reversal and domain structures in perpendicular synthetic antiferromagnets prepared on rigid and flexible substrates}
\author{Shaktiranjan Mohanty}
\affiliation{Laboratory for Nanomagnetism and Magnetic Materials (LNMM), School of Physical Sciences, National Institute of Science Education and Research (NISER),  An OCC of Homi Bhabha National Institute (HBNI), Jatni 752050, Odisha, India}

\author{Minaxi Sharma}
\altaffiliation{Present address: School of Basic Sciences, Bahra University, Waknaghat- 173234, Solan-H.P.}

\author{Ashish Kumar Moharana}
\affiliation{Laboratory for Nanomagnetism and Magnetic Materials (LNMM), School of Physical Sciences, National Institute of Science Education and Research (NISER),  An OCC of Homi Bhabha National Institute (HBNI), Jatni 752050, Odisha, India}

\author{Brindaban Ojha}
\affiliation{Laboratory for Nanomagnetism and Magnetic Materials (LNMM), School of Physical Sciences, National Institute of Science Education and Research (NISER),  An OCC of Homi Bhabha National Institute (HBNI), Jatni 752050, Odisha, India}

\author{Esita Pandey}
\affiliation{Laboratory for Nanomagnetism and Magnetic Materials (LNMM), School of Physical Sciences, National Institute of Science Education and Research (NISER),  An OCC of Homi Bhabha National Institute (HBNI), Jatni 752050, Odisha, India}

\author{Braj Bhusan Singh}
\affiliation{Laboratory for Nanomagnetism and Magnetic Materials (LNMM), School of Physical Sciences, National Institute of Science Education and Research (NISER),  An OCC of Homi Bhabha National Institute (HBNI), Jatni 752050, Odisha, India}

\author{Subhankar Bedanta}
\email{sbedanta@niser.ac.in}
\affiliation{Laboratory for Nanomagnetism and Magnetic Materials (LNMM), School of Physical Sciences, National Institute of Science Education and Research (NISER),  An OCC of Homi Bhabha National Institute (HBNI), Jatni 752050, Odisha, India}
\affiliation{Center for Interdisciplinary Sciences(CIS), National Institute of Science Education and Research (NISER),  An OCC of Homi Bhabha National Institute (HBNI), Jatni 752050, Odisha, India
}

\date{\today}

\begin{abstract}
Ferromagnetic (FM) layers separated by nonmagnetic metallic spacer layers can exhibit Ruderman–Kittel–Kasuya–Yosida (RKKY) coupling which may lead to a stable synthetic antiferromagnetic (SAF) phase. In this article we study magnetization reversal in [Co/Pt] layers by varying the number of bilayer stacks (Pt/Co) as well as thickness of Ir space layer t$_{Ir}$ on rigid Si(100) and flexible polyimide substrates. The samples with t$_{Ir}$ = 1.0 nm shows a FM coupling whereas samples with t$_{Ir}$ =  1.5 nm shows an AFM coupling between the FM layers. At t$_{Ir}$ = 2.0 nm, it shows a bow-tie shaped hysteresis loop indicating a canting of magnetization at the reversal. Higher anisotropy energy as compared to the interlayer exchange coupling (IEC) energy is an indication of the smaller relative angle between the magnetization of lower and upper FM layers. We have also demonstrated the strain induced modification of IEC as well as magnetization reversal phenomena. The IEC shows a slight decrease upon application of compressive strain and increase upon application of tensile strain which indicates the potential of SAFs in flexible spintronics. 
\end{abstract}

\maketitle

\section{\label{sec:level1}Introduction}
Tunability, miniaturization and functionality of spintronic devices depend on several factors including the interface engineering of ferromagnetic (FM) and nonmagnetic (NM) ultrathin layers. Recently, SAFs have drawn an immense research interest due to its flexibility over antiferromagnetic materials\cite{duine2018synthetic}. It consists of two FM layers separated by a metallic spacer layer, having antiparallel magnetization among the consecutive FM layers\cite{duine2018synthetic, bandiera2010comparison}. The FM layers are coupled by the IEC which is the RKKY type exchange coupling. This coupling is of oscillatory nature which oscillates with the thickness of NM layer. Therefore, in a FM/NM/FM system it is possible to observe global FM or AFM behaviour depending on the thickness of the spacer\cite{bruno1991oscillatory,stiles1999interlayer,bruno1995theory}. The coupling strength gradually decays with increasing thickness of the NM layer. SAF gives an extra degree of freedom over FM materials for avoiding the stray field. Further SAF offers the robustness against external perturbation which helps to tune stability and sensitivity of the devices. Magnetic tunnel junctions (MTJs) already play a pivotal role for enhancing the data storage capacity. SAFs are the key part of these MTJ devices for better data retaining and thermal stability due to the absence of stray field. Apart from these, there are several other aspects for which SAFs are in focus now a days for spintronic research. Recently, S.H. Yang et. al. has reported a higher domain wall velocity in Co/Ni layers with Ru spacer\cite{yang2015domain}. Q. Yang et. al. has shown that the RKKY interaction can be tuned by applying voltage\cite{yang2018ionic}. Also, SAF with manganite based LCMO (La$_{2/3}$/Ca$_{1/3}$MnO$_3$) as the oxide ferromagnetic layer and insulating spacer of CRTO (CaRu$_{1/2}$Ti$_{1/2}$O$_3$)  has been reported\cite{chen2017all}. Further SAFs became more promising now-a-days as it hosts chiral magnetic textures viz. skyrmions\cite{legrand2020room}.  IEC between two FM layers helps to minimize skyrmion Hall effect \cite{fert2013skyrmions, everschor2018perspective, chen2017skyrmion, dohi2019formation}. SAFs have been fabricated by using metallic spacer layers such as Ru, Ir, Rh etc. In order to use SAF in flexible spintronics it is of utmost importance to well understand their magnetization reversal phenomena with and without strain. In this context here, we have fabricated SAFs by taking Pt/Co as FM layer and Ir as spacer layer deposited on both rigid and flexible substrates.  For the SAFs with Ir thickness as 1.0 nm, 1.5 nm and 2.0 nm fabricated on rigid Si(100) substrates, we have observed three different types of spin configuration among the FM layers i.e., FM coupling, AFM coupling and canted magnetic configuration, respectively. Then the effect of such couplings on modifying the domain structure of SAF samples is also discussed. Further in order to study the effect of strain on the magnetization reversal phenomena of SAFs, we have prepared the SAF samples on flexible substrates. Flexible electronics have seen major technological innovation in recent times. The fabrication of electronic devices on flexible substrates has led to the development of bendable electronic skin and display\cite{mannsfeld2010highly, miyamoto2017inflammation}.Fabrication of the SAF with perpendicular magnetic anisotropy on flexible substrates require special attention since the quality of the interface between layers strongly depend upon the roughness of substrate used and the thickness of the buffer layer\cite{pandey2020strain, vemulkar2016toward}.  Hence preparing such thin films on flexible substrates such as polyimide is an important step to proceed further in this direction. The effect of stress on modifying the PMA and AFM exchange coupling between the FM layers is also studied which may be helpful for future flexible spintronic devices.

\section{Experimental details} 

\begin{table*}
\caption{\label{tab:table3}Sample name and structure}
\begin{ruledtabular}
\begin{tabular}{lll}

Sample name & Structure & (m, n) \\
\hline

R1 &    Si/SiO$_{2}$/Ta(3)/[Pt(3.5)/Co(0.8)]$_{2}$/Ta(3) &  \\

S1 &    Si/SiO$_{2}$/Ta(3)/[Pt(3.5)/Co(0.8)]$_{2}$/Ir(1.0)/[Co(0.8)/Pt(3.5)] &  2, 1\\

S2 &    Si/SiO$_{2}$/Ta(3)/[Pt(3.5)/Co(0.8)]$_{2}$/Ir(1.5)/[Co(0.8)/Pt(3.5)] &  2, 1\\

S3 &	Si/SiO$_{2}$/Ta(3)/[Pt(3.5)/Co(0.8)]$_{2}$/Ir(2.0)/[Co(0.8)/Pt(3.5)] &  2, 1\\

S4 &	Si/SiO$_{2}$/Ta(3)/[Pt(3.5)/Co(0.8)]/Ir(1.0)/[Co(0.8)/Pt(3.5)] &  1, 1\\

S5 &	Si/SiO$_{2}$/Ta(3)/[Pt(3.5)/Co(0.8)]/Ir(1.5)/[Co(0.8)/Pt(3.5)] &  1, 1\\

S6 &	Si/SiO$_{2}$/Ta(3)/[Pt(3.5)/Co(0.8)]/Ir(2.0)/[Co(0.8)/Pt(3.5)] &  1, 1\\

S7 &	Si/SiO$_{2}$/Ta(3)/[Pt(3.5)/Co(0.8)]/Ir(1.5)/[Co(0.8)/Pt(3.5)]$_{2}$ &  1, 2\\

S8 &	Si/SiO$_{2}$/Ta(3)/[Pt(3.5)/Co(0.8)]$_{2}$/Ir(1.5)/{[Co(0.8)/Pt(3.5)]}$_{2}$ &  2, 2\\

F1 &	PI/Ta(15)/[Pt(3.5)/Co(0.9])/Ir(1.5)/[Co(0.9)/Pt(3.5]) &  1, 1\\

\end{tabular}
\end{ruledtabular}
\end{table*}

A total of 9 samples have been prepared on rigid Si(100) substrates with sample structure Si/Ta(3)/[Pt(3.5)/Co(0.8)]$_{m}$/Ir($t_{Ir}$)/[Co(0.8)/Pt(3.5)]$_{n}$ along with a reference sample (R1) of Si/Ta(3)/[Pt(3.5)/Co(0.8)]$_{2}$/Ta(3). All the thicknesses shown in the parenthesis are in nm. In the sample structure, (m, n) are (2, 1) and (1, 1), respectively, for $t_{Ir}$ $=$ 1.0, 1.5, and 2.0 nm. We name these 6 samples as S1, S2, S3, S4, S5, S6. Another two samples are with $t_{Ir}$ = 1.5 nm and (m, n) are (1, 2) and (2, 2) which are named as S7 and S8. All the sample names and their structures are listed in table 1. Schematics of the sample structure is shown in figure 1(a). A seed layer Ta is grown on the substrate. It favors the growth of Pt layer in order to provide PMA in the Co layer. In all the samples, (except reference one), Pt is used as a capping layer in order to prevent oxidation of the Co layer.\\
In order to study the effect of stress on the strength of IEC for the magnetic multilayers, another one sample has been prepared on 38$\mu$m thick flexible polyimide substrate (manufactured by Dupont). The sample structure is PI/Ta(15)/Pt(3.5)/Co(0.9)/Ir($t_{Ir}$= 1.5)/Co(0.9)/Pt(3.5). All the thicknesses shown in the parenthesis are in nm. This sample is named as F1. Here, the Ta seed layer is taken as 15 nm to reduce the effect roughness of the polyimide substrate on the grown film.
All the samples have been prepared in a high vacuum multi-deposition chamber manufactured by Mantis Deposition Ltd., UK. The base pressure of the chamber was better than $1 \times 10^{-7}$ mbar. The deposition pressure was $\sim 1.5 \times 10^{-3}$ mbar for Ta and Pt layers. Further for Co and Ir, the deposition pressures were $\sim 5\times 10^{-3}$ mbar and $\sim 2.1\times 10^{-3}$ mbar, respectively. During sample preparation, the substrate table was rotated at 15 rpm in order to minimize the growth induced anisotropy and also to have uniform growth of the films. The rate of deposition were 0.1 \AA/s, 0.13 \AA/s, 0.3 \AA/s, 0.1 \AA/s for Ir, Ta, Pt and Co, respectively. \\
For the structural characterization of these ultrathin layers, we have performed cross-sectional TEM imaging in a high-resolution transmission electron microscope (HRTEM) (JEOL F200, operating at 200 kV and equipped with a GATAN oneview CMOS camera). For the cross sectional TEM sample preparation, two small pieces of samples were attached with film side facing each other by using epoxy glue. After several process of cutting, grinding and dimpling using diamond wire cutter, disc grinder and dimple grinder, the processed 3mm dia sample was milled with Ar ion in the PIPS (precision ion polishing system) II manufactured by GATAN. The milling was performed with an ion beam energy of 5 keV and a milling angle of 4$^{\circ}$ both below and above by two ion guns in the dual modulation mode.\\
Simultaneous observation of magnetic domain images and hysteresis loop in polar mode have been performed for all the samples with the help of magneto optic Kerr effect (MOKE) based microscope manufactured by Evico magnetics GmbH, Germany. For quantifying the interlayer exchange coupling energy and anisotropy energy etc. we have performed the hysteresis measurement at room temperature by a SQUID-VSM (Superconducting Quantum Interference Device-Vibrating sample magnetometer) manufactured by Quantum Design, USA.

\section{Result and Discussion}

\begin{figure}[h]
	\centering
	\includegraphics[width=0.48\textwidth]{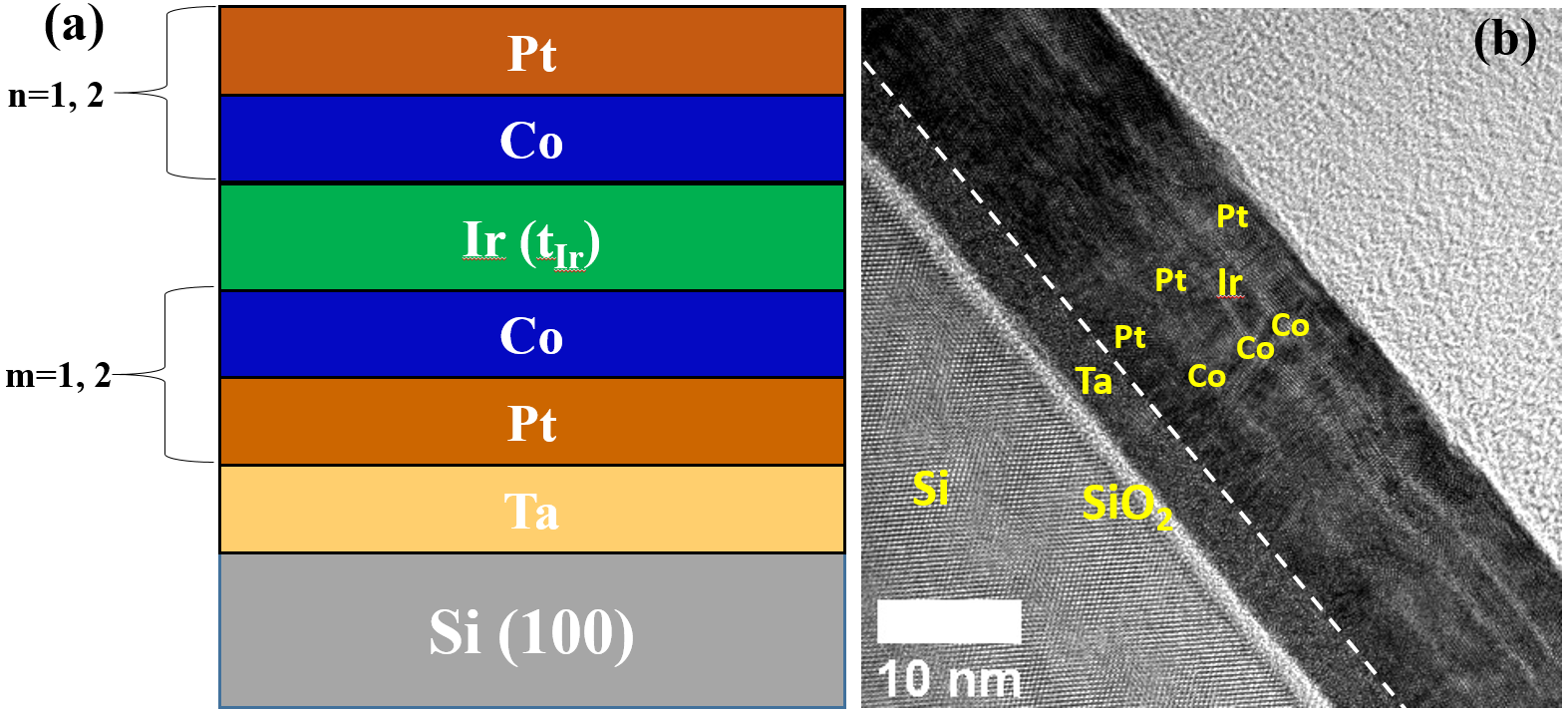}
	\caption{\label{fig:fig1}{(a) Schematic of the sample structure.(b) Cross-sectional TEM image of the sample S2.}}
\end{figure}

Fig.\ref{fig:fig1}(b) shows the high resolution TEM images of the sample S2 mentioned above. The growth of individual layers is clearly visible (indicated in the image with respective layer names).

\begin{figure}[h]
	\centering
	\includegraphics[width=0.48\textwidth]{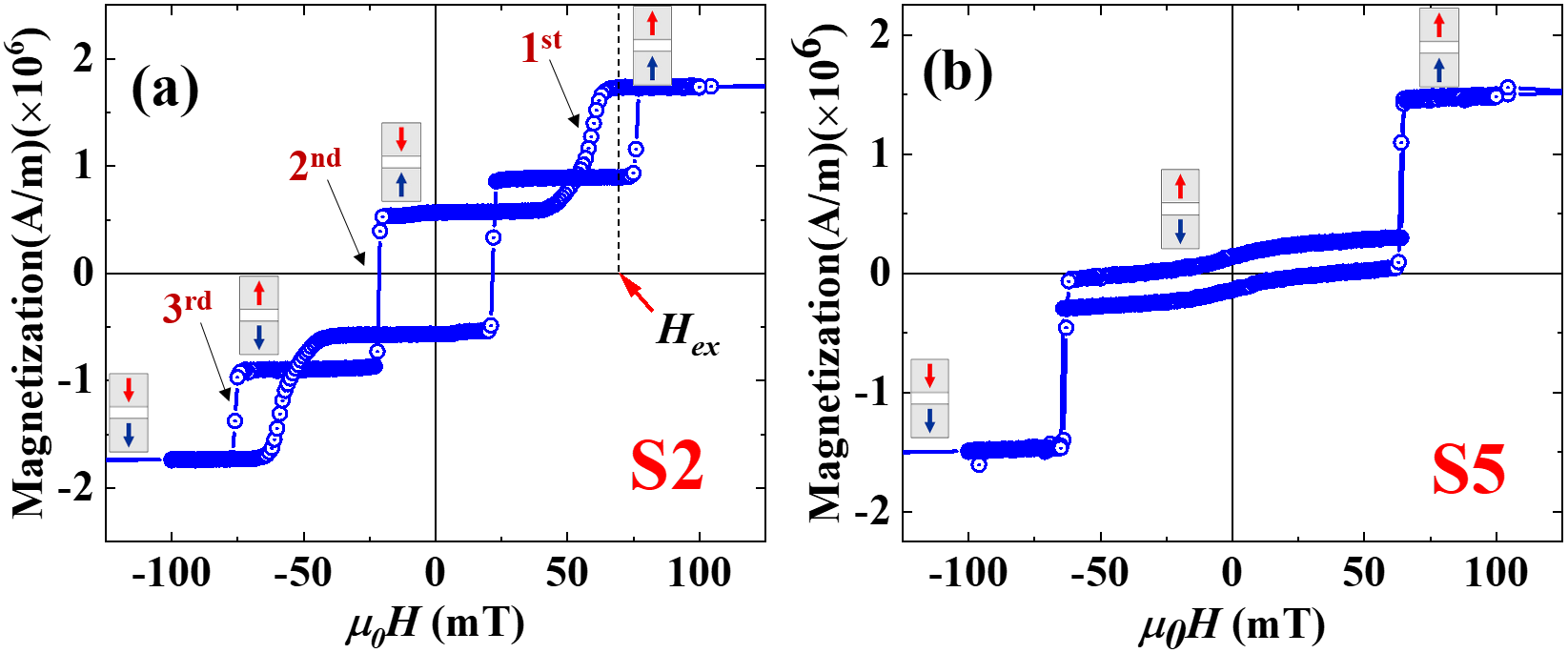}
	\caption{Magnetization reversal measured by SQUID-VSM for samples (a) S2 and (b) S5 where the magnetic field was applied perpendicular to the film plane.}
	\label{fig:fig2}
\end{figure}

  Hysteresis loop of the reference sample R1 measured by SQUID-VSM in presence of a perpendicular magnetic field has been shown in the Fig. S1 of supplementary information.
  Magnetic hysteresis loops of the samples S2 and S5 are shown in Fig.\ref{fig:fig2} (a) and (b), respectively, measured by SQUID-VSM. These two samples are with thickness of Ir spacer layer t$_{Ir}$=1.5nm with (m, n) as (2,1) and (1,1), respectively. The steps in the hysteresis loops indicate AFM coupling between the FM layers below and above the Ir spacer layer.  Fig.\ref{fig:fig2}(a) shows that the magnetization reversal is accompanied by three steps in the hysteresis loop. The magnetization reversal may be explained in the following manner. In the 1st reversal the top FM layer switches first and becomes AF coupled to the bottom FM layer. In the 2nd reversal, both the layers switch oppositely due to strong AFM coupling. Finally, under sufficiently negative external field (Zeeman energy) the top layer switches along the negative field direction and hence saturation is achieved.   For sample S5, there are only two reversals indicating that in the 1st reversal, both layers become AF coupled and in the 2nd reversal, the coupling is lost and both FM layers become negatively saturated. Further we observe in Fig. \ref{fig:fig2}(a) that there is a substantial remanent magnetization in the sample indicating that S2 is an uncompensated SAF but in fig. \ref{fig:fig2}(b), the remanence is almost close to zero which indicates that S5 is a compensated SAF. The other two samples S7 and S8 with t$_{Ir}$=1.5 nm and (m,n) values (1,2) and (2,2), respectively, also show the AFM coupling between the FM layers which is shown in the supplementary information. The IEC energy of these AF coupled FM layers can be calculated by using the expression J$_{ex}$=H$_{ex}$M$_{s}$t where, H$_{ex}$ is the exchange coupling field at which the coupling between the FM layer is vanishes and the magnetization directions in both the layers become parallel\cite{yakushiji2017very, gabor2017interlayer}. M$_{s}$ is the saturation magnetization and t is the thickness of the FM layers\cite{yang2018ionic}. H$_{ex}$ can be calculated form the hysteresis loop measured by SQUID-VSM as indicated in the Fig.\ref{fig:fig2}(a). The IEC energy (J$_{ex}$) of AF coupled samples S2, S5, S7 and S8 are found to be 2.96$\times$$10^{-4}$$J/m^2$, 1.64 $\times$$10^{-4}$$J/m^2$, 2.74 $\times$$10^{-4}$$J/m^2$ and 4.56 $\times$$10^{-4}$$J/m^2$, respectively. Here it shows an increase in the coupling strength with the increase in the number of Co/Pt layers indicating that it needs more energy to break the coupling for more number of Co/Pt layers below and above the spacer. The samples S1 and S4 with t$_{Ir}$=1.0 nm, shows FM coupling indicating that this thickness of Ir is not in the AFM coupling regime. M-H loops corresponding to these samples are shown in supplementary information. The other two samples S3 and S6 with t$_{Ir}$=2.0 nm show a bow-tie shape hysteresis loop with no steps. It indicates that the AF coupling is reduced at this Ir thickness. 
  
  \begin{figure}[h]
	\centering
	\includegraphics[width=0.48\textwidth]{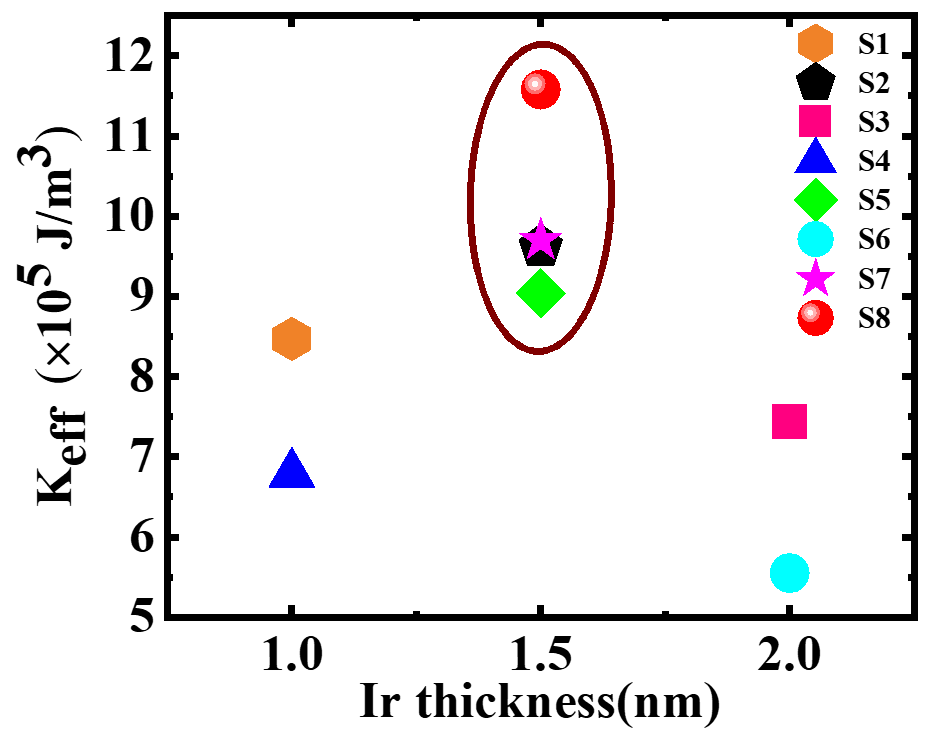}
	\caption{Effective anisotropy energy density calculated for the samples S1 to S8}
	\label{fig:fig3}
\end{figure}

  The effective anisotropy energy of all the samples is evaluated by measuring the hysteresis loops along both in-plane and out-of-plane directions of the samples in SQUID magnetometer. In this context, we have used the equation K$_{eff}$ = $\mu_0$H$_{k}$M$_{s}$/2, where $\mu_0$H$_{k}$ is the anisotropy field (in-plane saturation field) and M$_s$ is the saturation magnetization (refer to table S2 in supplementary information). We found an enhancement in the anisotropy energy of the samples with AFM coupling as compared to the samples with FM coupling. The anisotropy energies of these samples are shown in Fig.\ref{fig:fig3} where The points inside the ellipse shows the effective anisotropy energy values for the SAF samples. The anisotropy energy of the samples are higher than the IEC energy (J$_{ex}$/t)  resulting in a smaller intermediate angle between the magnetic moments of the two layers\cite{yakushiji2015perpendicular, liu2019strong}.\\

\begin{figure*}
	\centering
	\includegraphics[width=1\textwidth]{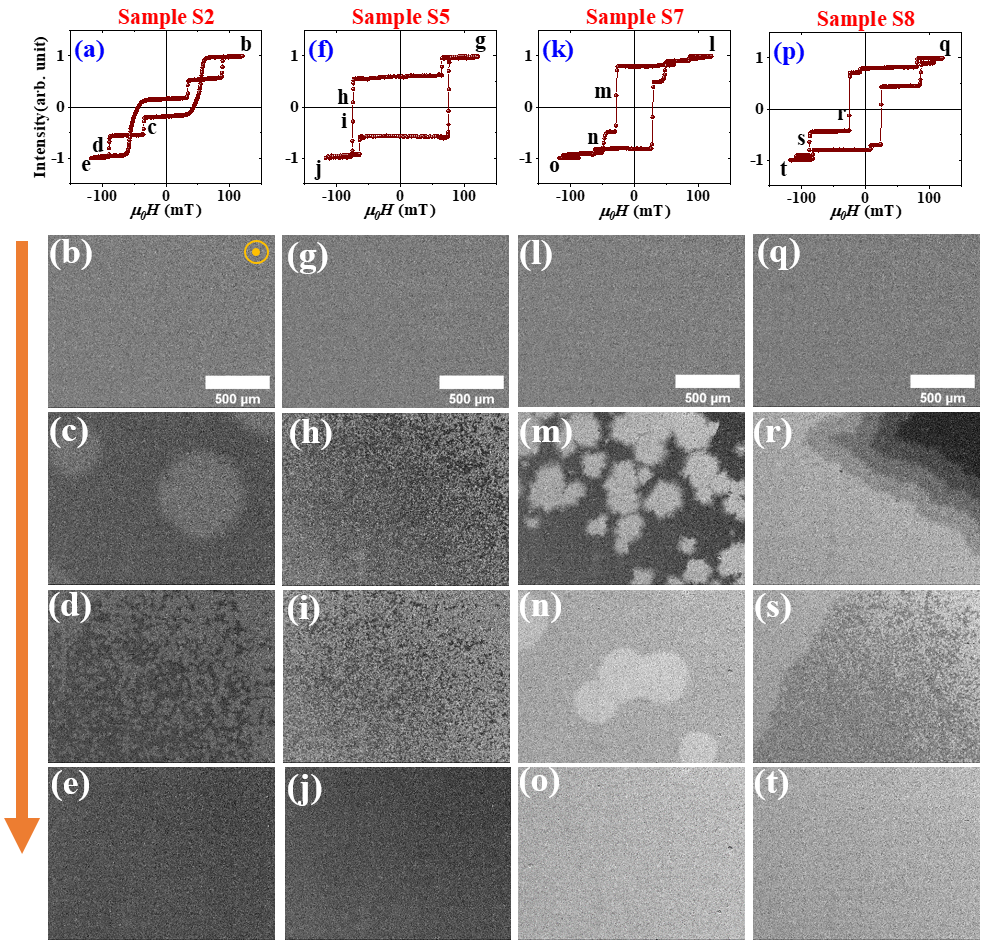}
\caption{\label{fig:fig4}Hysteresis loops and domain images of the SAF samples S2, S5, S7 and S8 measured.}
\end{figure*}

Fig.\ref{fig:fig4} shows the hysteresis loops and the corresponding domain images measured by MOKE microscopy in polar mode for the AFM coupled samples S2, S5, S7 and  S8. From left to right, the hysteresis loops represent the samples S2, S5, S7, S8, respectively, and below each loop the domain images are shown from top to bottom at respective points mentioned in the loop.

For sample S2 the hysteresis has 3 steps as shown in Fig.\ref{fig:fig4}.(a). Here the magnetization reversal is achieved by both domain nucleation and domain wall (DW) propagation (Fig.\ref{fig:fig4}. (b-e)). Here we have defined the magnetization directions for the Pt/Co layers above and below the Ir spacer layer by red arrow ($\color{red}\uparrow$) and blue arrow ($\color{blue}\uparrow$), respectively. In the first reversal ($\color{red}\uparrow \color{blue}\uparrow - \color{red}\downarrow \color{blue}\uparrow$) no domain has been observed and only a change in the contrast of domain image is there. Such type of domain behavior may be explained as the spin flop transition which can be inferred from the slanted reversal of the hysteresis loop at the first reversal\cite{yun2021tuning, talantsev2018relaxation}.
However in the second reversal ($\color{red}\downarrow \color{blue}\uparrow - \color{red}\uparrow \color{blue}\downarrow$) bubble domains have been observed due to the sharp transition (Fig.\ref{fig:fig4}.c). In the third reversal ($\color{red}\uparrow \color{blue}\downarrow - \color{red}\downarrow \color{blue}\downarrow$) the size of the domains became remarkably small which may be due to the lowering of AFM coupling between the layers Fig.\ref{fig:fig4} (d)\cite{fu2009tuning}. In sample S5 only two step reversal (see Fig.\ref{fig:fig4}(f)) has been achieved via a large nucleation of small bubble domains (Fig.\ref{fig:fig4}(h-i)).
\begin{figure}[h]
	\centering
	\includegraphics[width=0.48\textwidth]{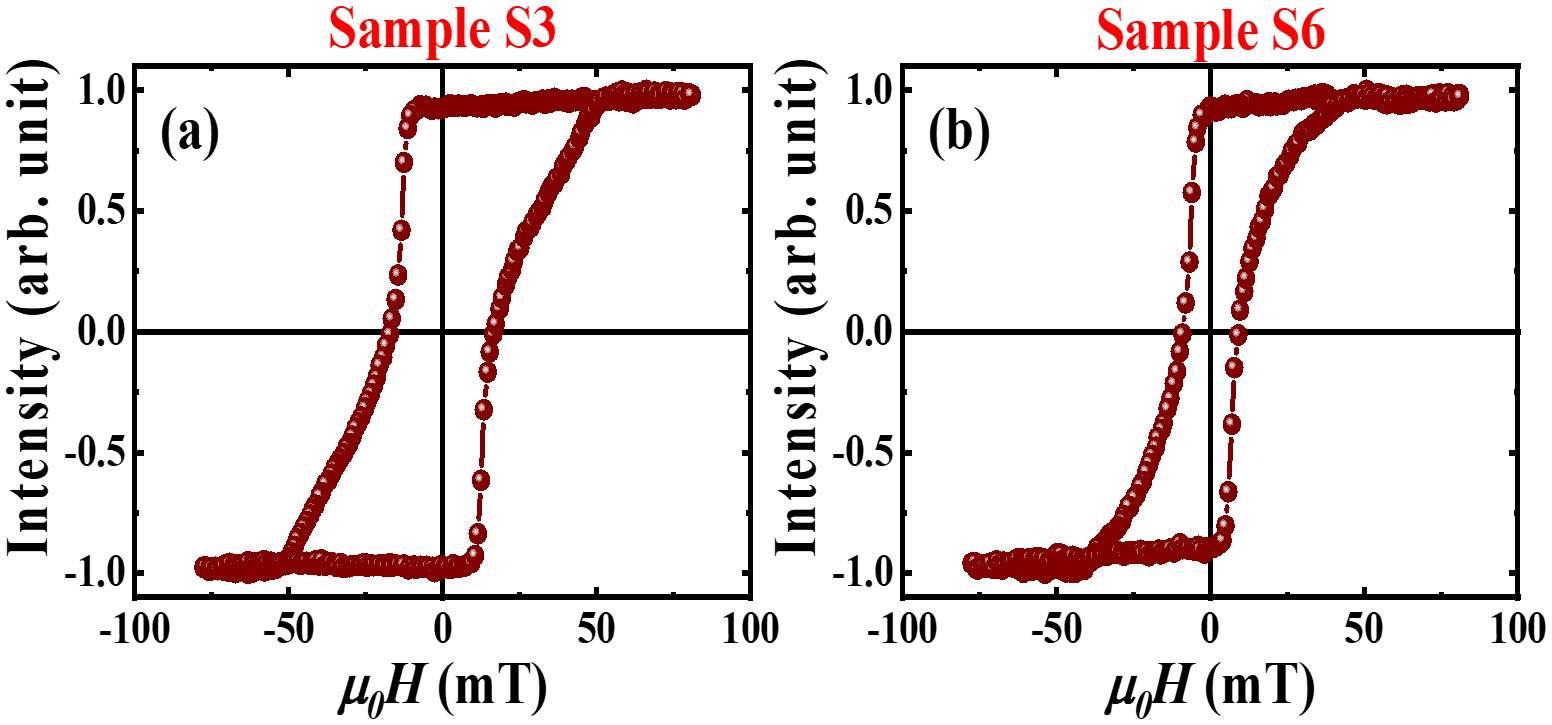}
	\caption{Hysteresis loops measured by magneto-optic Kerr effect based microscopy in polar mode for samples S3 and S6.}
	\label{fig:fig5}
\end{figure}
\begin{figure}[h]
	\centering
	\includegraphics[width=0.48\textwidth]{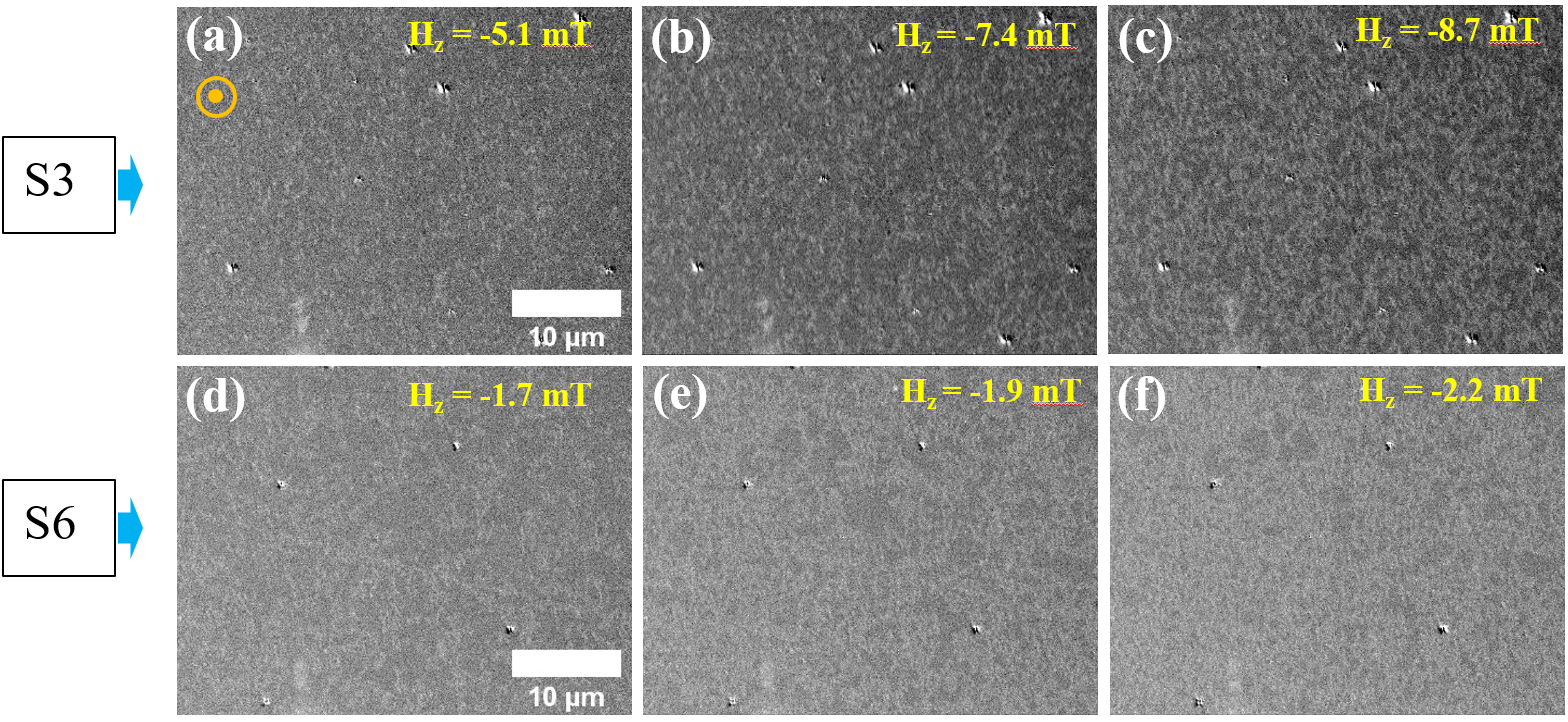}
	\caption{ Domain images of the samples S3 (a - c) and S6 (d - f) at the reversal. Scale bar shown in (a) and (d) are valid for all other images. }
	\label{fig:fig6}
\end{figure}
In the sample S7, a three step hysteresis loop is observed which is shown in Fig.\ref{fig:fig4} (k). Here, two different types of domains are observed at two different reversal i.e. distorted bubble domain with more number of nucleation and symmetric bubble domain as shown in (m) and (n), respectively. Here the bubble domains are quite similar to sample S2 due to their comparable IEC strength. Sample S8 exhibits a multi-step magnetization reversal where in the first step only single domain propagation has been observed. In the second reversal also a single domain formation and its propagation occurred where apart from dark and bright contrast the other intermediate contrasts indicate a continuous rotation of the magnetization vector in the sample. In the last reversal two different kind of domains (one big domain and other small bubble domains) simultaneously propagated during the reversal and finally saturated along the field direction.
The samples S3 and S6 show a bow-tie shaped hysteresis loop measured by magneto optic Kerr effect microscopy shown in Fig. \ref{fig:fig5}. At the reversal, very small ripple kind of domains are observed which is indicated in the Fig.\ref{fig:fig6}. The hysteresis loops for sample S3 and S6 shown in Fig. \ref{fig:fig5} have been measured using a 5X objective however due to very small domain size, the domain images were captured using a 50X objective (as shown in Fig. \ref{fig:fig6}).
\begin{figure}[h]
	\centering
	\includegraphics[width=0.48\textwidth]{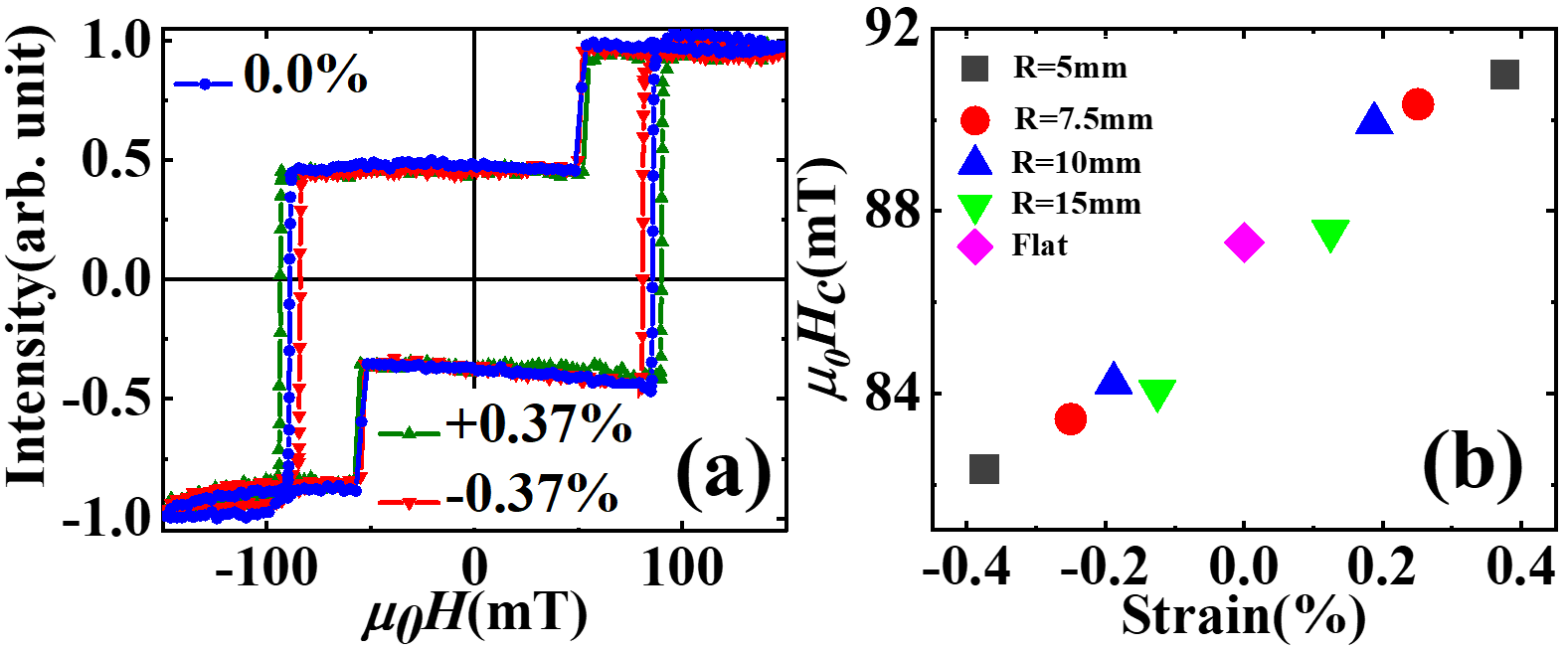}
	\caption{Hysteresis plot of Co/Pt synthetic antiferromagnet with Ir thickness of 1.5 nm under 0.37$\%$ of tensile and compressive strain. (b) Plot of coercive field by varying the magnitude of strain. }
	\label{fig:fig7}
\end{figure}
In order to understand the effect of strain on the magnetization reversal, henceforth we will discuss the results on sample prepared on flexible PI substrate. 
Fig.\ref{fig:fig7} shows the Kerr microscopy hysteresis loops of SAF prepared on polyimide substrates at both strained and unstrained state. In addition, the hysteresis loop of its Si counterpart is also measured (see supplementary information) where sharp switching fields are observed for samples prepared on both the substrates. Comparable coercive fields for the same sample on both the substrates indicate that the strength of the magnetic anisotropy and RKKY coupling is quite identical which makes flexible SAFs very promising from application viewpoint.
To generate tensile and compressive strains the flexible sample has been fixed on both convex and concave shaped molds respectively. The magnitude of strain is varied by using molds of different radius of curvature \cite{pandey2020strain}. The magnitude of strain generated in this manner can be calculated by the following equation \cite{tang2014magneto, qiao2017enhanced, dai2012mechanically, wang2005magnetostriction}:

\begin{equation}
    \epsilon_{\pm}= \frac{t_{total}}{2R\pm t_{total}}
\end{equation}

where $\epsilon_+$  denotes the applied tensile strain and $\epsilon_-$ is for compressive strain, R is the radius of the mold used and $t_{total}$ is the total thickness of the magnetic thin film and flexible substrate. \\
Fig.\ref{fig:fig7}(a) shows the hysteresis loop of sample F1 under both flat and bend states. It shows a slight change in coercivity under both compressive and tensile stress. Under 0.37$\%$ tensile stress, we observe an increase in coercivity of 3.09 mT while the compressive stress decreases the coercivity by 3.85 mT. Hence only a slight modification ($\sim$4\%) in coercivity has been observed for the sample under a moderate strain ($\sim$0.4\%).  The change in coercivity is attributed to the change in PMA as a result of the introduction of stress induced anisotropy\cite{cullity2011introduction}. The stress induced anisotropy is given by

\begin{equation}
    E_\sigma=\frac{3}{2}\lambda_s \sigma sin^2(\theta)
\end{equation}
 
where $\lambda_s$ is magnetostriction coefficient, $\sigma$ is the applied stress, $\theta$ is the angle between the $\sigma$ and magnetization vector. It has been shown in earlier reports that the $\lambda$ is negative for Co \cite{shepley2015modification}. \\

For a negative magnetostrictive FM material the stress induced anisotropy acts along the in-plane direction for compressive strain and out-of-plane direction for tensile strain. 
As a result, the stress induced anisotropy reduces the PMA of SAF under compressive stress and enhances the anisotropy in the case of tensile stress. Fig.\ref{fig:fig7}(b) shows the variation in coercive field by varying the magnitude of both type of strain. With the increase in compressive strain, there is a systematic decrease in coercivity while with tensile strain the coercivity increases continuously. Such changes in coercivity under strain can be attributed to a change in PMA because of magneto-elastic anisotropy \cite{vemulkar2016toward, shepley2015modification, bairagi2018experimental}. The interlayer exchange coupling field ($\mu_{0}$H$_{ex}$) has also changed under the effect of strain which is a clear indication of the change in coupling energy. Exchange coupling field for flat state is 69.88 mT. With the effect of tensile strain, the $\mu_{0}$H$_{ex}$ increases to 73.47 mT and decreases to 67.54 mT in case of the compressive strain. However, no significant changes of domain images has been observed under application of both type of strain (shown in supplementary information).

\section{Conclusion}
In this work SAF samples with perpendicular magnetic anisotropy have been prepared on both rigid and flexible substrates. The variation of iridium spacer layer thickness leads to different types of coupling among the FM layers with different spin configurations i.e., FM and AFM couplings. Comparison of IEC energy and anisotropy energy suggests the relatively smaller angle between the spins of FM layers below and above the spacer layer in SAF samples. On the other hand, the effect of strain has shown a substantial change of the exchange coupling field of the SAF sample which leads to the change of interlayer exchange coupling energy. The exchange coupling field shows an increase upon application of tensile strain and decrease upon application of compressive strain. We further noticed that the application of strain does not lead to any significant change of domain structure in the SAF sample. Further work is required to understand the detailed reversal processes in such SAFs. 

\section{Supplementary information}
The supplementary information is available with the manuscript.

\section{acknowledgments}
We acknowledge the financial support by the Department of Atomic Energy (DAE), Govt. of India for providing the financial support. We also thank the CEFIPRA project for their funding. BBS acknowledges DST for INSPIRE faculty fellowship.

\section{References}

\bibliography{aipsamp}

\providecommand{\noopsort}[1]{}\providecommand{\singleletter}[1]{#1}%
\begin{thebibliography}{31}%
\makeatletter
\providecommand \@ifxundefined [1]{%
 \@ifx{#1\undefined}
}%
\providecommand \@ifnum [1]{%
 \ifnum #1\expandafter \@firstoftwo
 \else \expandafter \@secondoftwo
 \fi
}%
\providecommand \@ifx [1]{%
 \ifx #1\expandafter \@firstoftwo
 \else \expandafter \@secondoftwo
 \fi
}%
\providecommand \natexlab [1]{#1}%
\providecommand \enquote  [1]{``#1''}%
\providecommand \bibnamefont  [1]{#1}%
\providecommand \bibfnamefont [1]{#1}%
\providecommand \citenamefont [1]{#1}%
\providecommand \href@noop [0]{\@secondoftwo}%
\providecommand \href [0]{\begingroup \@sanitize@url \@href}%
\providecommand \@href[1]{\@@startlink{#1}\@@href}%
\providecommand \@@href[1]{\endgroup#1\@@endlink}%
\providecommand \@sanitize@url [0]{\catcode `\\12\catcode `\$12\catcode
  `\&12\catcode `\#12\catcode `\^12\catcode `\_12\catcode `\%12\relax}%
\providecommand \@@startlink[1]{}%
\providecommand \@@endlink[0]{}%
\providecommand \url  [0]{\begingroup\@sanitize@url \@url }%
\providecommand \@url [1]{\endgroup\@href {#1}{\urlprefix }}%
\providecommand \urlprefix  [0]{URL }%
\providecommand \Eprint [0]{\href }%
\providecommand \doibase [0]{https://doi.org/}%
\providecommand \selectlanguage [0]{\@gobble}%
\providecommand \bibinfo  [0]{\@secondoftwo}%
\providecommand \bibfield  [0]{\@secondoftwo}%
\providecommand \translation [1]{[#1]}%
\providecommand \BibitemOpen [0]{}%
\providecommand \bibitemStop [0]{}%
\providecommand \bibitemNoStop [0]{.\EOS\space}%
\providecommand \EOS [0]{\spacefactor3000\relax}%
\providecommand \BibitemShut  [1]{\csname bibitem#1\endcsname}%
\let\auto@bib@innerbib\@empty
\bibitem [{\citenamefont {Duine}\ \emph {et~al.}(2018)\citenamefont {Duine},
  \citenamefont {Lee}, \citenamefont {Parkin},\ and\ \citenamefont
  {Stiles}}]{duine2018synthetic}%
  \BibitemOpen
  \bibfield  {author} {\bibinfo {author} {\bibfnamefont {R.}~\bibnamefont
  {Duine}}, \bibinfo {author} {\bibfnamefont {K.-J.}\ \bibnamefont {Lee}},
  \bibinfo {author} {\bibfnamefont {S.~S.}\ \bibnamefont {Parkin}},\ and\
  \bibinfo {author} {\bibfnamefont {M.~D.}\ \bibnamefont {Stiles}},\ }\bibfield
   {title} {\enquote {\bibinfo {title} {Synthetic antiferromagnetic
  spintronics},}\ }\href@noop {} {\bibfield  {journal} {\bibinfo  {journal}
  {Nature physics}\ }\textbf {\bibinfo {volume} {14}},\ \bibinfo {pages}
  {217--219} (\bibinfo {year} {2018})}\BibitemShut {NoStop}%
\bibitem [{\citenamefont {Bandiera}\ \emph {et~al.}(2010)\citenamefont
  {Bandiera}, \citenamefont {Sousa}, \citenamefont {Dahmane}, \citenamefont
  {Ducruet}, \citenamefont {Portemont}, \citenamefont {Baltz}, \citenamefont
  {Auffret}, \citenamefont {Prejbeanu},\ and\ \citenamefont
  {Dieny}}]{bandiera2010comparison}%
  \BibitemOpen
  \bibfield  {author} {\bibinfo {author} {\bibfnamefont {S.}~\bibnamefont
  {Bandiera}}, \bibinfo {author} {\bibfnamefont {R.~C.}\ \bibnamefont {Sousa}},
  \bibinfo {author} {\bibfnamefont {Y.}~\bibnamefont {Dahmane}}, \bibinfo
  {author} {\bibfnamefont {C.}~\bibnamefont {Ducruet}}, \bibinfo {author}
  {\bibfnamefont {C.}~\bibnamefont {Portemont}}, \bibinfo {author}
  {\bibfnamefont {V.}~\bibnamefont {Baltz}}, \bibinfo {author} {\bibfnamefont
  {S.}~\bibnamefont {Auffret}}, \bibinfo {author} {\bibfnamefont {I.~L.}\
  \bibnamefont {Prejbeanu}},\ and\ \bibinfo {author} {\bibfnamefont
  {B.}~\bibnamefont {Dieny}},\ }\bibfield  {title} {\enquote {\bibinfo {title}
  {Comparison of synthetic antiferromagnets and hard ferromagnets as reference
  layer in magnetic tunnel junctions with perpendicular magnetic anisotropy},}\
  }\href@noop {} {\bibfield  {journal} {\bibinfo  {journal} {IEEE Magnetics
  Letters}\ }\textbf {\bibinfo {volume} {1}},\ \bibinfo {pages}
  {3000204--3000204} (\bibinfo {year} {2010})}\BibitemShut {NoStop}%
\bibitem [{\citenamefont {Bruno}\ and\ \citenamefont
  {Chappert}(1991)}]{bruno1991oscillatory}%
  \BibitemOpen
  \bibfield  {author} {\bibinfo {author} {\bibfnamefont {e.~P.}\ \bibnamefont
  {Bruno}}\ and\ \bibinfo {author} {\bibfnamefont {C.}~\bibnamefont
  {Chappert}},\ }\bibfield  {title} {\enquote {\bibinfo {title} {Oscillatory
  coupling between ferromagnetic layers separated by a nonmagnetic metal
  spacer},}\ }\href@noop {} {\bibfield  {journal} {\bibinfo  {journal}
  {Physical review letters}\ }\textbf {\bibinfo {volume} {67}},\ \bibinfo
  {pages} {1602} (\bibinfo {year} {1991})}\BibitemShut {NoStop}%
\bibitem [{\citenamefont {Stiles}(1999)}]{stiles1999interlayer}%
  \BibitemOpen
  \bibfield  {author} {\bibinfo {author} {\bibfnamefont {M.~D.}\ \bibnamefont
  {Stiles}},\ }\bibfield  {title} {\enquote {\bibinfo {title} {Interlayer
  exchange coupling},}\ }\href@noop {} {\bibfield  {journal} {\bibinfo
  {journal} {Journal of Magnetism and Magnetic Materials}\ }\textbf {\bibinfo
  {volume} {200}},\ \bibinfo {pages} {322--337} (\bibinfo {year}
  {1999})}\BibitemShut {NoStop}%
\bibitem [{\citenamefont {Bruno}(1995)}]{bruno1995theory}%
  \BibitemOpen
  \bibfield  {author} {\bibinfo {author} {\bibfnamefont {P.}~\bibnamefont
  {Bruno}},\ }\bibfield  {title} {\enquote {\bibinfo {title} {Theory of
  interlayer magnetic coupling},}\ }\href@noop {} {\bibfield  {journal}
  {\bibinfo  {journal} {Physical Review B}\ }\textbf {\bibinfo {volume} {52}},\
  \bibinfo {pages} {411} (\bibinfo {year} {1995})}\BibitemShut {NoStop}%
\bibitem [{\citenamefont {Yang}, \citenamefont {Ryu},\ and\ \citenamefont
  {Parkin}(2015)}]{yang2015domain}%
  \BibitemOpen
  \bibfield  {author} {\bibinfo {author} {\bibfnamefont {S.-H.}\ \bibnamefont
  {Yang}}, \bibinfo {author} {\bibfnamefont {K.-S.}\ \bibnamefont {Ryu}},\ and\
  \bibinfo {author} {\bibfnamefont {S.}~\bibnamefont {Parkin}},\ }\bibfield
  {title} {\enquote {\bibinfo {title} {Domain-wall velocities of up to 750 m s-
  1 driven by exchange-coupling torque in synthetic antiferromagnets},}\
  }\href@noop {} {\bibfield  {journal} {\bibinfo  {journal} {Nature
  nanotechnology}\ }\textbf {\bibinfo {volume} {10}},\ \bibinfo {pages}
  {221--226} (\bibinfo {year} {2015})}\BibitemShut {NoStop}%
\bibitem [{\citenamefont {Yang}\ \emph {et~al.}(2018)\citenamefont {Yang},
  \citenamefont {Wang}, \citenamefont {Zhou}, \citenamefont {Wang},
  \citenamefont {Zhang}, \citenamefont {Zhao}, \citenamefont {Dong},
  \citenamefont {Cheng}, \citenamefont {Min}, \citenamefont {Hu} \emph
  {et~al.}}]{yang2018ionic}%
  \BibitemOpen
  \bibfield  {author} {\bibinfo {author} {\bibfnamefont {Q.}~\bibnamefont
  {Yang}}, \bibinfo {author} {\bibfnamefont {L.}~\bibnamefont {Wang}}, \bibinfo
  {author} {\bibfnamefont {Z.}~\bibnamefont {Zhou}}, \bibinfo {author}
  {\bibfnamefont {L.}~\bibnamefont {Wang}}, \bibinfo {author} {\bibfnamefont
  {Y.}~\bibnamefont {Zhang}}, \bibinfo {author} {\bibfnamefont
  {S.}~\bibnamefont {Zhao}}, \bibinfo {author} {\bibfnamefont {G.}~\bibnamefont
  {Dong}}, \bibinfo {author} {\bibfnamefont {Y.}~\bibnamefont {Cheng}},
  \bibinfo {author} {\bibfnamefont {T.}~\bibnamefont {Min}}, \bibinfo {author}
  {\bibfnamefont {Z.}~\bibnamefont {Hu}}, \emph {et~al.},\ }\bibfield  {title}
  {\enquote {\bibinfo {title} {Ionic liquid gating control of rkky interaction
  in fecob/ru/fecob and (pt/co) 2/ru/(co/pt) 2 multilayers},}\ }\href@noop {}
  {\bibfield  {journal} {\bibinfo  {journal} {Nature communications}\ }\textbf
  {\bibinfo {volume} {9}},\ \bibinfo {pages} {1--11} (\bibinfo {year}
  {2018})}\BibitemShut {NoStop}%
\bibitem [{\citenamefont {Chen}\ \emph {et~al.}(2017)\citenamefont {Chen},
  \citenamefont {Xu}, \citenamefont {Ma}, \citenamefont {Mattauch},
  \citenamefont {Lan}, \citenamefont {Jin}, \citenamefont {Guo}, \citenamefont
  {Wan}, \citenamefont {Chen}, \citenamefont {Gao} \emph
  {et~al.}}]{chen2017all}%
  \BibitemOpen
  \bibfield  {author} {\bibinfo {author} {\bibfnamefont {B.}~\bibnamefont
  {Chen}}, \bibinfo {author} {\bibfnamefont {H.}~\bibnamefont {Xu}}, \bibinfo
  {author} {\bibfnamefont {C.}~\bibnamefont {Ma}}, \bibinfo {author}
  {\bibfnamefont {S.}~\bibnamefont {Mattauch}}, \bibinfo {author}
  {\bibfnamefont {D.}~\bibnamefont {Lan}}, \bibinfo {author} {\bibfnamefont
  {F.}~\bibnamefont {Jin}}, \bibinfo {author} {\bibfnamefont {Z.}~\bibnamefont
  {Guo}}, \bibinfo {author} {\bibfnamefont {S.}~\bibnamefont {Wan}}, \bibinfo
  {author} {\bibfnamefont {P.}~\bibnamefont {Chen}}, \bibinfo {author}
  {\bibfnamefont {G.}~\bibnamefont {Gao}}, \emph {et~al.},\ }\bibfield  {title}
  {\enquote {\bibinfo {title} {All-oxide--based synthetic antiferromagnets
  exhibiting layer-resolved magnetization reversal},}\ }\href@noop {}
  {\bibfield  {journal} {\bibinfo  {journal} {Science}\ }\textbf {\bibinfo
  {volume} {357}},\ \bibinfo {pages} {191--194} (\bibinfo {year}
  {2017})}\BibitemShut {NoStop}%
\bibitem [{\citenamefont {Legrand}\ \emph {et~al.}(2020)\citenamefont
  {Legrand}, \citenamefont {Maccariello}, \citenamefont {Ajejas}, \citenamefont
  {Collin}, \citenamefont {Vecchiola}, \citenamefont {Bouzehouane},
  \citenamefont {Reyren}, \citenamefont {Cros},\ and\ \citenamefont
  {Fert}}]{legrand2020room}%
  \BibitemOpen
  \bibfield  {author} {\bibinfo {author} {\bibfnamefont {W.}~\bibnamefont
  {Legrand}}, \bibinfo {author} {\bibfnamefont {D.}~\bibnamefont
  {Maccariello}}, \bibinfo {author} {\bibfnamefont {F.}~\bibnamefont {Ajejas}},
  \bibinfo {author} {\bibfnamefont {S.}~\bibnamefont {Collin}}, \bibinfo
  {author} {\bibfnamefont {A.}~\bibnamefont {Vecchiola}}, \bibinfo {author}
  {\bibfnamefont {K.}~\bibnamefont {Bouzehouane}}, \bibinfo {author}
  {\bibfnamefont {N.}~\bibnamefont {Reyren}}, \bibinfo {author} {\bibfnamefont
  {V.}~\bibnamefont {Cros}},\ and\ \bibinfo {author} {\bibfnamefont
  {A.}~\bibnamefont {Fert}},\ }\bibfield  {title} {\enquote {\bibinfo {title}
  {Room-temperature stabilization of antiferromagnetic skyrmions in synthetic
  antiferromagnets},}\ }\href@noop {} {\bibfield  {journal} {\bibinfo
  {journal} {Nature materials}\ }\textbf {\bibinfo {volume} {19}},\ \bibinfo
  {pages} {34--42} (\bibinfo {year} {2020})}\BibitemShut {NoStop}%
\bibitem [{\citenamefont {Fert}, \citenamefont {Cros},\ and\ \citenamefont
  {Sampaio}(2013)}]{fert2013skyrmions}%
  \BibitemOpen
  \bibfield  {author} {\bibinfo {author} {\bibfnamefont {A.}~\bibnamefont
  {Fert}}, \bibinfo {author} {\bibfnamefont {V.}~\bibnamefont {Cros}},\ and\
  \bibinfo {author} {\bibfnamefont {J.}~\bibnamefont {Sampaio}},\ }\bibfield
  {title} {\enquote {\bibinfo {title} {Skyrmions on the track},}\ }\href@noop
  {} {\bibfield  {journal} {\bibinfo  {journal} {Nature nanotechnology}\
  }\textbf {\bibinfo {volume} {8}},\ \bibinfo {pages} {152--156} (\bibinfo
  {year} {2013})}\BibitemShut {NoStop}%
\bibitem [{\citenamefont {Everschor-Sitte}\ \emph {et~al.}(2018)\citenamefont
  {Everschor-Sitte}, \citenamefont {Masell}, \citenamefont {Reeve},\ and\
  \citenamefont {Kl{\"a}ui}}]{everschor2018perspective}%
  \BibitemOpen
  \bibfield  {author} {\bibinfo {author} {\bibfnamefont {K.}~\bibnamefont
  {Everschor-Sitte}}, \bibinfo {author} {\bibfnamefont {J.}~\bibnamefont
  {Masell}}, \bibinfo {author} {\bibfnamefont {R.~M.}\ \bibnamefont {Reeve}},\
  and\ \bibinfo {author} {\bibfnamefont {M.}~\bibnamefont {Kl{\"a}ui}},\
  }\bibfield  {title} {\enquote {\bibinfo {title} {Perspective: Magnetic
  skyrmions—overview of recent progress in an active research field},}\
  }\href@noop {} {\bibfield  {journal} {\bibinfo  {journal} {Journal of Applied
  Physics}\ }\textbf {\bibinfo {volume} {124}},\ \bibinfo {pages} {240901}
  (\bibinfo {year} {2018})}\BibitemShut {NoStop}%
\bibitem [{\citenamefont {Chen}(2017)}]{chen2017skyrmion}%
  \BibitemOpen
  \bibfield  {author} {\bibinfo {author} {\bibfnamefont {G.}~\bibnamefont
  {Chen}},\ }\bibfield  {title} {\enquote {\bibinfo {title} {Skyrmion hall
  effect},}\ }\href@noop {} {\bibfield  {journal} {\bibinfo  {journal} {Nature
  Physics}\ }\textbf {\bibinfo {volume} {13}},\ \bibinfo {pages} {112--113}
  (\bibinfo {year} {2017})}\BibitemShut {NoStop}%
\bibitem [{\citenamefont {Dohi}\ \emph {et~al.}(2019)\citenamefont {Dohi},
  \citenamefont {DuttaGupta}, \citenamefont {Fukami},\ and\ \citenamefont
  {Ohno}}]{dohi2019formation}%
  \BibitemOpen
  \bibfield  {author} {\bibinfo {author} {\bibfnamefont {T.}~\bibnamefont
  {Dohi}}, \bibinfo {author} {\bibfnamefont {S.}~\bibnamefont {DuttaGupta}},
  \bibinfo {author} {\bibfnamefont {S.}~\bibnamefont {Fukami}},\ and\ \bibinfo
  {author} {\bibfnamefont {H.}~\bibnamefont {Ohno}},\ }\bibfield  {title}
  {\enquote {\bibinfo {title} {Formation and current-induced motion of
  synthetic antiferromagnetic skyrmion bubbles},}\ }\href@noop {} {\bibfield
  {journal} {\bibinfo  {journal} {Nature communications}\ }\textbf {\bibinfo
  {volume} {10}},\ \bibinfo {pages} {1--6} (\bibinfo {year}
  {2019})}\BibitemShut {NoStop}%
\bibitem [{\citenamefont {Mannsfeld}\ \emph {et~al.}(2010)\citenamefont
  {Mannsfeld}, \citenamefont {Tee}, \citenamefont {Stoltenberg}, \citenamefont
  {Chen}, \citenamefont {Barman}, \citenamefont {Muir}, \citenamefont
  {Sokolov}, \citenamefont {Reese},\ and\ \citenamefont
  {Bao}}]{mannsfeld2010highly}%
  \BibitemOpen
  \bibfield  {author} {\bibinfo {author} {\bibfnamefont {S.~C.}\ \bibnamefont
  {Mannsfeld}}, \bibinfo {author} {\bibfnamefont {B.~C.}\ \bibnamefont {Tee}},
  \bibinfo {author} {\bibfnamefont {R.~M.}\ \bibnamefont {Stoltenberg}},
  \bibinfo {author} {\bibfnamefont {C.~V.}\ \bibnamefont {Chen}}, \bibinfo
  {author} {\bibfnamefont {S.}~\bibnamefont {Barman}}, \bibinfo {author}
  {\bibfnamefont {B.~V.}\ \bibnamefont {Muir}}, \bibinfo {author}
  {\bibfnamefont {A.~N.}\ \bibnamefont {Sokolov}}, \bibinfo {author}
  {\bibfnamefont {C.}~\bibnamefont {Reese}},\ and\ \bibinfo {author}
  {\bibfnamefont {Z.}~\bibnamefont {Bao}},\ }\bibfield  {title} {\enquote
  {\bibinfo {title} {Highly sensitive flexible pressure sensors with
  microstructured rubber dielectric layers},}\ }\href@noop {} {\bibfield
  {journal} {\bibinfo  {journal} {Nature materials}\ }\textbf {\bibinfo
  {volume} {9}},\ \bibinfo {pages} {859--864} (\bibinfo {year}
  {2010})}\BibitemShut {NoStop}%
\bibitem [{\citenamefont {Miyamoto}\ \emph {et~al.}(2017)\citenamefont
  {Miyamoto}, \citenamefont {Lee}, \citenamefont {Cooray}, \citenamefont {Lee},
  \citenamefont {Mori}, \citenamefont {Matsuhisa}, \citenamefont {Jin},
  \citenamefont {Yoda}, \citenamefont {Yokota}, \citenamefont {Itoh} \emph
  {et~al.}}]{miyamoto2017inflammation}%
  \BibitemOpen
  \bibfield  {author} {\bibinfo {author} {\bibfnamefont {A.}~\bibnamefont
  {Miyamoto}}, \bibinfo {author} {\bibfnamefont {S.}~\bibnamefont {Lee}},
  \bibinfo {author} {\bibfnamefont {N.~F.}\ \bibnamefont {Cooray}}, \bibinfo
  {author} {\bibfnamefont {S.}~\bibnamefont {Lee}}, \bibinfo {author}
  {\bibfnamefont {M.}~\bibnamefont {Mori}}, \bibinfo {author} {\bibfnamefont
  {N.}~\bibnamefont {Matsuhisa}}, \bibinfo {author} {\bibfnamefont
  {H.}~\bibnamefont {Jin}}, \bibinfo {author} {\bibfnamefont {L.}~\bibnamefont
  {Yoda}}, \bibinfo {author} {\bibfnamefont {T.}~\bibnamefont {Yokota}},
  \bibinfo {author} {\bibfnamefont {A.}~\bibnamefont {Itoh}}, \emph {et~al.},\
  }\bibfield  {title} {\enquote {\bibinfo {title} {Inflammation-free,
  gas-permeable, lightweight, stretchable on-skin electronics with
  nanomeshes},}\ }\href@noop {} {\bibfield  {journal} {\bibinfo  {journal}
  {Nature nanotechnology}\ }\textbf {\bibinfo {volume} {12}},\ \bibinfo {pages}
  {907--913} (\bibinfo {year} {2017})}\BibitemShut {NoStop}%
\bibitem [{\citenamefont {Pandey}\ \emph {et~al.}(2020)\citenamefont {Pandey},
  \citenamefont {Singh}, \citenamefont {Sharangi},\ and\ \citenamefont
  {Bedanta}}]{pandey2020strain}%
  \BibitemOpen
  \bibfield  {author} {\bibinfo {author} {\bibfnamefont {E.}~\bibnamefont
  {Pandey}}, \bibinfo {author} {\bibfnamefont {B.~B.}\ \bibnamefont {Singh}},
  \bibinfo {author} {\bibfnamefont {P.}~\bibnamefont {Sharangi}},\ and\
  \bibinfo {author} {\bibfnamefont {S.}~\bibnamefont {Bedanta}},\ }\bibfield
  {title} {\enquote {\bibinfo {title} {Strain engineered domain structure and
  their relaxation in perpendicularly magnetized co/pt deposited on flexible
  polyimide},}\ }\href@noop {} {\bibfield  {journal} {\bibinfo  {journal} {Nano
  Express}\ }\textbf {\bibinfo {volume} {1}},\ \bibinfo {pages} {010037}
  (\bibinfo {year} {2020})}\BibitemShut {NoStop}%
\bibitem [{\citenamefont {Vemulkar}\ \emph {et~al.}(2016)\citenamefont
  {Vemulkar}, \citenamefont {Mansell}, \citenamefont {Fern{\'a}ndez-Pacheco},\
  and\ \citenamefont {Cowburn}}]{vemulkar2016toward}%
  \BibitemOpen
  \bibfield  {author} {\bibinfo {author} {\bibfnamefont {T.}~\bibnamefont
  {Vemulkar}}, \bibinfo {author} {\bibfnamefont {R.}~\bibnamefont {Mansell}},
  \bibinfo {author} {\bibfnamefont {A.}~\bibnamefont {Fern{\'a}ndez-Pacheco}},\
  and\ \bibinfo {author} {\bibfnamefont {R.}~\bibnamefont {Cowburn}},\
  }\bibfield  {title} {\enquote {\bibinfo {title} {Toward flexible spintronics:
  Perpendicularly magnetized synthetic antiferromagnetic thin films and
  nanowires on polyimide substrates},}\ }\href@noop {} {\bibfield  {journal}
  {\bibinfo  {journal} {Advanced Functional Materials}\ }\textbf {\bibinfo
  {volume} {26}},\ \bibinfo {pages} {4704--4711} (\bibinfo {year}
  {2016})}\BibitemShut {NoStop}%
\bibitem [{\citenamefont {Yakushiji}\ \emph {et~al.}(2017)\citenamefont
  {Yakushiji}, \citenamefont {Sugihara}, \citenamefont {Fukushima},
  \citenamefont {Kubota},\ and\ \citenamefont {Yuasa}}]{yakushiji2017very}%
  \BibitemOpen
  \bibfield  {author} {\bibinfo {author} {\bibfnamefont {K.}~\bibnamefont
  {Yakushiji}}, \bibinfo {author} {\bibfnamefont {A.}~\bibnamefont {Sugihara}},
  \bibinfo {author} {\bibfnamefont {A.}~\bibnamefont {Fukushima}}, \bibinfo
  {author} {\bibfnamefont {H.}~\bibnamefont {Kubota}},\ and\ \bibinfo {author}
  {\bibfnamefont {S.}~\bibnamefont {Yuasa}},\ }\bibfield  {title} {\enquote
  {\bibinfo {title} {Very strong antiferromagnetic interlayer exchange coupling
  with iridium spacer layer for perpendicular magnetic tunnel junctions},}\
  }\href@noop {} {\bibfield  {journal} {\bibinfo  {journal} {Applied Physics
  Letters}\ }\textbf {\bibinfo {volume} {110}},\ \bibinfo {pages} {092406}
  (\bibinfo {year} {2017})}\BibitemShut {NoStop}%
\bibitem [{\citenamefont {Gabor}\ \emph {et~al.}(2017)\citenamefont {Gabor},
  \citenamefont {Petrisor}, \citenamefont {Mos}, \citenamefont {Nasui},\ and\
  \citenamefont {Tiusan}}]{gabor2017interlayer}%
  \BibitemOpen
  \bibfield  {author} {\bibinfo {author} {\bibfnamefont {M.}~\bibnamefont
  {Gabor}}, \bibinfo {author} {\bibfnamefont {T.}~\bibnamefont {Petrisor}},
  \bibinfo {author} {\bibfnamefont {R.}~\bibnamefont {Mos}}, \bibinfo {author}
  {\bibfnamefont {M.}~\bibnamefont {Nasui}},\ and\ \bibinfo {author}
  {\bibfnamefont {C.}~\bibnamefont {Tiusan}},\ }\bibfield  {title} {\enquote
  {\bibinfo {title} {Interlayer exchange coupling in perpendicularly magnetized
  pt/co/ir/co/pt structures},}\ }\href@noop {} {\bibfield  {journal} {\bibinfo
  {journal} {Journal of Physics D: Applied Physics}\ }\textbf {\bibinfo
  {volume} {50}},\ \bibinfo {pages} {465004} (\bibinfo {year}
  {2017})}\BibitemShut {NoStop}%
\bibitem [{\citenamefont {Yakushiji}\ \emph {et~al.}(2015)\citenamefont
  {Yakushiji}, \citenamefont {Kubota}, \citenamefont {Fukushima},\ and\
  \citenamefont {Yuasa}}]{yakushiji2015perpendicular}%
  \BibitemOpen
  \bibfield  {author} {\bibinfo {author} {\bibfnamefont {K.}~\bibnamefont
  {Yakushiji}}, \bibinfo {author} {\bibfnamefont {H.}~\bibnamefont {Kubota}},
  \bibinfo {author} {\bibfnamefont {A.}~\bibnamefont {Fukushima}},\ and\
  \bibinfo {author} {\bibfnamefont {S.}~\bibnamefont {Yuasa}},\ }\bibfield
  {title} {\enquote {\bibinfo {title} {Perpendicular magnetic tunnel junctions
  with strong antiferromagnetic interlayer exchange coupling at first
  oscillation peak},}\ }\href@noop {} {\bibfield  {journal} {\bibinfo
  {journal} {Applied Physics Express}\ }\textbf {\bibinfo {volume} {8}},\
  \bibinfo {pages} {083003} (\bibinfo {year} {2015})}\BibitemShut {NoStop}%
\bibitem [{\citenamefont {Liu}, \citenamefont {Yu},\ and\ \citenamefont
  {Zhong}(2019)}]{liu2019strong}%
  \BibitemOpen
  \bibfield  {author} {\bibinfo {author} {\bibfnamefont {Y.}~\bibnamefont
  {Liu}}, \bibinfo {author} {\bibfnamefont {J.}~\bibnamefont {Yu}},\ and\
  \bibinfo {author} {\bibfnamefont {H.}~\bibnamefont {Zhong}},\ }\bibfield
  {title} {\enquote {\bibinfo {title} {Strong antiferromagnetic interlayer
  exchange coupling in [co/pt] 6/ru/[co/pt] 4 structures with perpendicular
  magnetic anisotropy},}\ }\href@noop {} {\bibfield  {journal} {\bibinfo
  {journal} {Journal of Magnetism and Magnetic Materials}\ }\textbf {\bibinfo
  {volume} {473}},\ \bibinfo {pages} {381--386} (\bibinfo {year}
  {2019})}\BibitemShut {NoStop}%
\bibitem [{\citenamefont {Yun}\ \emph {et~al.}(2021)\citenamefont {Yun},
  \citenamefont {Sheng}, \citenamefont {Guo}, \citenamefont {Zheng},
  \citenamefont {Chen}, \citenamefont {Cao}, \citenamefont {Yan}, \citenamefont
  {He}, \citenamefont {Jin}, \citenamefont {Li} \emph
  {et~al.}}]{yun2021tuning}%
  \BibitemOpen
  \bibfield  {author} {\bibinfo {author} {\bibfnamefont {J.}~\bibnamefont
  {Yun}}, \bibinfo {author} {\bibfnamefont {Y.}~\bibnamefont {Sheng}}, \bibinfo
  {author} {\bibfnamefont {X.}~\bibnamefont {Guo}}, \bibinfo {author}
  {\bibfnamefont {B.}~\bibnamefont {Zheng}}, \bibinfo {author} {\bibfnamefont
  {P.}~\bibnamefont {Chen}}, \bibinfo {author} {\bibfnamefont {Y.}~\bibnamefont
  {Cao}}, \bibinfo {author} {\bibfnamefont {Z.}~\bibnamefont {Yan}}, \bibinfo
  {author} {\bibfnamefont {X.}~\bibnamefont {He}}, \bibinfo {author}
  {\bibfnamefont {P.}~\bibnamefont {Jin}}, \bibinfo {author} {\bibfnamefont
  {J.}~\bibnamefont {Li}}, \emph {et~al.},\ }\bibfield  {title} {\enquote
  {\bibinfo {title} {Tuning the spin-flop transition in perpendicularly
  magnetized synthetic antiferromagnets by swift heavy fe ions irradiation},}\
  }\href@noop {} {\bibfield  {journal} {\bibinfo  {journal} {Physical Review
  B}\ }\textbf {\bibinfo {volume} {104}},\ \bibinfo {pages} {134416} (\bibinfo
  {year} {2021})}\BibitemShut {NoStop}%
\bibitem [{\citenamefont {Talantsev}\ \emph {et~al.}(2018)\citenamefont
  {Talantsev}, \citenamefont {Lu}, \citenamefont {Fache}, \citenamefont
  {Lavanant}, \citenamefont {Hamadeh}, \citenamefont {Aristov}, \citenamefont
  {Koplak}, \citenamefont {Morgunov},\ and\ \citenamefont
  {Mangin}}]{talantsev2018relaxation}%
  \BibitemOpen
  \bibfield  {author} {\bibinfo {author} {\bibfnamefont {A.}~\bibnamefont
  {Talantsev}}, \bibinfo {author} {\bibfnamefont {Y.}~\bibnamefont {Lu}},
  \bibinfo {author} {\bibfnamefont {T.}~\bibnamefont {Fache}}, \bibinfo
  {author} {\bibfnamefont {M.}~\bibnamefont {Lavanant}}, \bibinfo {author}
  {\bibfnamefont {A.}~\bibnamefont {Hamadeh}}, \bibinfo {author} {\bibfnamefont
  {A.}~\bibnamefont {Aristov}}, \bibinfo {author} {\bibfnamefont
  {O.}~\bibnamefont {Koplak}}, \bibinfo {author} {\bibfnamefont
  {R.}~\bibnamefont {Morgunov}},\ and\ \bibinfo {author} {\bibfnamefont
  {S.}~\bibnamefont {Mangin}},\ }\bibfield  {title} {\enquote {\bibinfo {title}
  {Relaxation dynamics of magnetization transitions in synthetic
  antiferromagnet with perpendicular anisotropy},}\ }\href@noop {} {\bibfield
  {journal} {\bibinfo  {journal} {Journal of Physics: Condensed Matter}\
  }\textbf {\bibinfo {volume} {30}},\ \bibinfo {pages} {135804} (\bibinfo
  {year} {2018})}\BibitemShut {NoStop}%
\bibitem [{\citenamefont {Fu}\ \emph {et~al.}(2009)\citenamefont {Fu},
  \citenamefont {Ishio}, \citenamefont {Wang}, \citenamefont {Pei},
  \citenamefont {Hasegawa}, \citenamefont {Yamane},\ and\ \citenamefont
  {Saito}}]{fu2009tuning}%
  \BibitemOpen
  \bibfield  {author} {\bibinfo {author} {\bibfnamefont {Y.}~\bibnamefont
  {Fu}}, \bibinfo {author} {\bibfnamefont {S.}~\bibnamefont {Ishio}}, \bibinfo
  {author} {\bibfnamefont {T.}~\bibnamefont {Wang}}, \bibinfo {author}
  {\bibfnamefont {W.}~\bibnamefont {Pei}}, \bibinfo {author} {\bibfnamefont
  {T.}~\bibnamefont {Hasegawa}}, \bibinfo {author} {\bibfnamefont
  {H.}~\bibnamefont {Yamane}},\ and\ \bibinfo {author} {\bibfnamefont
  {H.}~\bibnamefont {Saito}},\ }\bibfield  {title} {\enquote {\bibinfo {title}
  {Tuning interlayer coupling and domain structure in [co/pd] n co/ru [co/pd
  (x)] m multilayer by controlling the thickness x of the weak-ferromagnetic pd
  layers in the lower co/pd multilayer},}\ }\href@noop {} {\bibfield  {journal}
  {\bibinfo  {journal} {Journal of Applied Physics}\ }\textbf {\bibinfo
  {volume} {105}},\ \bibinfo {pages} {07C307} (\bibinfo {year}
  {2009})}\BibitemShut {NoStop}%
\bibitem [{\citenamefont {Tang}\ \emph {et~al.}(2014)\citenamefont {Tang},
  \citenamefont {Wang}, \citenamefont {Yang}, \citenamefont {Xu}, \citenamefont
  {Liu}, \citenamefont {Sun}, \citenamefont {Xia}, \citenamefont {Zhan},
  \citenamefont {Chen}, \citenamefont {Tang} \emph {et~al.}}]{tang2014magneto}%
  \BibitemOpen
  \bibfield  {author} {\bibinfo {author} {\bibfnamefont {Z.}~\bibnamefont
  {Tang}}, \bibinfo {author} {\bibfnamefont {B.}~\bibnamefont {Wang}}, \bibinfo
  {author} {\bibfnamefont {H.}~\bibnamefont {Yang}}, \bibinfo {author}
  {\bibfnamefont {X.}~\bibnamefont {Xu}}, \bibinfo {author} {\bibfnamefont
  {Y.}~\bibnamefont {Liu}}, \bibinfo {author} {\bibfnamefont {D.}~\bibnamefont
  {Sun}}, \bibinfo {author} {\bibfnamefont {L.}~\bibnamefont {Xia}}, \bibinfo
  {author} {\bibfnamefont {Q.}~\bibnamefont {Zhan}}, \bibinfo {author}
  {\bibfnamefont {B.}~\bibnamefont {Chen}}, \bibinfo {author} {\bibfnamefont
  {M.}~\bibnamefont {Tang}}, \emph {et~al.},\ }\bibfield  {title} {\enquote
  {\bibinfo {title} {Magneto-mechanical coupling effect in amorphous
  co40fe40b20 films grown on flexible substrates},}\ }\href@noop {} {\bibfield
  {journal} {\bibinfo  {journal} {Applied Physics Letters}\ }\textbf {\bibinfo
  {volume} {105}},\ \bibinfo {pages} {103504} (\bibinfo {year}
  {2014})}\BibitemShut {NoStop}%
\bibitem [{\citenamefont {Qiao}\ \emph {et~al.}(2017)\citenamefont {Qiao},
  \citenamefont {Wen}, \citenamefont {Wang}, \citenamefont {Bai}, \citenamefont
  {Zhan}, \citenamefont {Xu},\ and\ \citenamefont {Li}}]{qiao2017enhanced}%
  \BibitemOpen
  \bibfield  {author} {\bibinfo {author} {\bibfnamefont {X.}~\bibnamefont
  {Qiao}}, \bibinfo {author} {\bibfnamefont {X.}~\bibnamefont {Wen}}, \bibinfo
  {author} {\bibfnamefont {B.}~\bibnamefont {Wang}}, \bibinfo {author}
  {\bibfnamefont {Y.}~\bibnamefont {Bai}}, \bibinfo {author} {\bibfnamefont
  {Q.}~\bibnamefont {Zhan}}, \bibinfo {author} {\bibfnamefont {X.}~\bibnamefont
  {Xu}},\ and\ \bibinfo {author} {\bibfnamefont {R.-W.}\ \bibnamefont {Li}},\
  }\bibfield  {title} {\enquote {\bibinfo {title} {Enhanced stress-invariance
  of magnetization direction in magnetic thin films},}\ }\href@noop {}
  {\bibfield  {journal} {\bibinfo  {journal} {Applied Physics Letters}\
  }\textbf {\bibinfo {volume} {111}},\ \bibinfo {pages} {132405} (\bibinfo
  {year} {2017})}\BibitemShut {NoStop}%
\bibitem [{\citenamefont {Dai}\ \emph {et~al.}(2012)\citenamefont {Dai},
  \citenamefont {Zhan}, \citenamefont {Liu}, \citenamefont {Yang},
  \citenamefont {Zhang}, \citenamefont {Chen},\ and\ \citenamefont
  {Li}}]{dai2012mechanically}%
  \BibitemOpen
  \bibfield  {author} {\bibinfo {author} {\bibfnamefont {G.}~\bibnamefont
  {Dai}}, \bibinfo {author} {\bibfnamefont {Q.}~\bibnamefont {Zhan}}, \bibinfo
  {author} {\bibfnamefont {Y.}~\bibnamefont {Liu}}, \bibinfo {author}
  {\bibfnamefont {H.}~\bibnamefont {Yang}}, \bibinfo {author} {\bibfnamefont
  {X.}~\bibnamefont {Zhang}}, \bibinfo {author} {\bibfnamefont
  {B.}~\bibnamefont {Chen}},\ and\ \bibinfo {author} {\bibfnamefont {R.-W.}\
  \bibnamefont {Li}},\ }\bibfield  {title} {\enquote {\bibinfo {title}
  {Mechanically tunable magnetic properties of fe81ga19 films grown on flexible
  substrates},}\ }\href@noop {} {\bibfield  {journal} {\bibinfo  {journal}
  {Applied Physics Letters}\ }\textbf {\bibinfo {volume} {100}},\ \bibinfo
  {pages} {122407} (\bibinfo {year} {2012})}\BibitemShut {NoStop}%
\bibitem [{\citenamefont {Wang}\ \emph {et~al.}(2005)\citenamefont {Wang},
  \citenamefont {Nordman}, \citenamefont {Qian}, \citenamefont {Daughton},\
  and\ \citenamefont {Myers}}]{wang2005magnetostriction}%
  \BibitemOpen
  \bibfield  {author} {\bibinfo {author} {\bibfnamefont {D.}~\bibnamefont
  {Wang}}, \bibinfo {author} {\bibfnamefont {C.}~\bibnamefont {Nordman}},
  \bibinfo {author} {\bibfnamefont {Z.}~\bibnamefont {Qian}}, \bibinfo {author}
  {\bibfnamefont {J.~M.}\ \bibnamefont {Daughton}},\ and\ \bibinfo {author}
  {\bibfnamefont {J.}~\bibnamefont {Myers}},\ }\bibfield  {title} {\enquote
  {\bibinfo {title} {Magnetostriction effect of amorphous cofeb thin films and
  application in spin-dependent tunnel junctions},}\ }\href@noop {} {\bibfield
  {journal} {\bibinfo  {journal} {Journal of applied physics}\ }\textbf
  {\bibinfo {volume} {97}},\ \bibinfo {pages} {10C906} (\bibinfo {year}
  {2005})}\BibitemShut {NoStop}%
\bibitem [{\citenamefont {Cullity}\ and\ \citenamefont
  {Graham}(2011)}]{cullity2011introduction}%
  \BibitemOpen
  \bibfield  {author} {\bibinfo {author} {\bibfnamefont {B.~D.}\ \bibnamefont
  {Cullity}}\ and\ \bibinfo {author} {\bibfnamefont {C.~D.}\ \bibnamefont
  {Graham}},\ }\href@noop {} {\emph {\bibinfo {title} {Introduction to magnetic
  materials}}}\ (\bibinfo  {publisher} {John Wiley \& Sons},\ \bibinfo {year}
  {2011})\BibitemShut {NoStop}%
\bibitem [{\citenamefont {Shepley}\ \emph {et~al.}(2015)\citenamefont
  {Shepley}, \citenamefont {Rushforth}, \citenamefont {Wang}, \citenamefont
  {Burnell},\ and\ \citenamefont {Moore}}]{shepley2015modification}%
  \BibitemOpen
  \bibfield  {author} {\bibinfo {author} {\bibfnamefont {P.}~\bibnamefont
  {Shepley}}, \bibinfo {author} {\bibfnamefont {A.}~\bibnamefont {Rushforth}},
  \bibinfo {author} {\bibfnamefont {M.}~\bibnamefont {Wang}}, \bibinfo {author}
  {\bibfnamefont {G.}~\bibnamefont {Burnell}},\ and\ \bibinfo {author}
  {\bibfnamefont {T.}~\bibnamefont {Moore}},\ }\bibfield  {title} {\enquote
  {\bibinfo {title} {Modification of perpendicular magnetic anisotropy and
  domain wall velocity in pt/co/pt by voltage-induced strain},}\ }\href@noop {}
  {\bibfield  {journal} {\bibinfo  {journal} {Scientific reports}\ }\textbf
  {\bibinfo {volume} {5}},\ \bibinfo {pages} {1--5} (\bibinfo {year}
  {2015})}\BibitemShut {NoStop}%
\bibitem [{\citenamefont {Bairagi}\ \emph {et~al.}(2018)\citenamefont
  {Bairagi}, \citenamefont {Bellec}, \citenamefont {Repain}, \citenamefont
  {Fourmental}, \citenamefont {Chacon}, \citenamefont {Girard}, \citenamefont
  {Lagoute}, \citenamefont {Rousset}, \citenamefont {Le~Laurent}, \citenamefont
  {Smogunov} \emph {et~al.}}]{bairagi2018experimental}%
  \BibitemOpen
  \bibfield  {author} {\bibinfo {author} {\bibfnamefont {K.}~\bibnamefont
  {Bairagi}}, \bibinfo {author} {\bibfnamefont {A.}~\bibnamefont {Bellec}},
  \bibinfo {author} {\bibfnamefont {V.}~\bibnamefont {Repain}}, \bibinfo
  {author} {\bibfnamefont {C.}~\bibnamefont {Fourmental}}, \bibinfo {author}
  {\bibfnamefont {C.}~\bibnamefont {Chacon}}, \bibinfo {author} {\bibfnamefont
  {Y.}~\bibnamefont {Girard}}, \bibinfo {author} {\bibfnamefont
  {J.}~\bibnamefont {Lagoute}}, \bibinfo {author} {\bibfnamefont
  {S.}~\bibnamefont {Rousset}}, \bibinfo {author} {\bibfnamefont
  {L.}~\bibnamefont {Le~Laurent}}, \bibinfo {author} {\bibfnamefont
  {A.}~\bibnamefont {Smogunov}}, \emph {et~al.},\ }\bibfield  {title} {\enquote
  {\bibinfo {title} {Experimental and theoretical investigations of magnetic
  anisotropy and magnetic hardening at molecule/ferromagnet interfaces},}\
  }\href@noop {} {\bibfield  {journal} {\bibinfo  {journal} {Physical Review
  B}\ }\textbf {\bibinfo {volume} {98}},\ \bibinfo {pages} {085432} (\bibinfo
  {year} {2018})}\BibitemShut {NoStop}%
\end{thebibliography}%

\end{document}